\documentstyle[twocolumn,prl,floats,aps,epsfig]{revtex}

\begin{document}

\draft

\wideabs{
\title{Magnetoresistance in Heavily Underdoped YBa$_2$Cu$_3$O$_{6+x}$:
Antiferromagnetic Correlations and Normal-State Transport}

\author{A. N. Lavrov\cite{ANL}, Yoichi Ando, Kouji Segawa and J. Takeya}

\address{Central Research Institute of Electric Power Industry,
2-11-1 Iwato-kita, Komae, Tokyo 201-8511, Japan}

\date{\today}

\maketitle

\begin{abstract}
We report on a contrasting behavior of the in-plane and out-of-plane
magnetoresistance (MR) in heavily underdoped antiferromagnetic (AF)
YBa$_2$Cu$_3$O$_{6+x}$ ($x$$\leq$$0.37$). The out-of-plane MR
($I$$\parallel$$c$) is positive over most of the temperature range and
shows a sharp increase, by about two orders of magnitude, upon cooling
through the N\'{e}el temperature $T_N$. A contribution associated with the
AF correlations is found to dominate the out-of-plane MR behavior for
$H$$\parallel$$c$ from far above $T_N$, pointing to the key role of spin
fluctuations in the out-of-plane transport. In contrast, the transverse
in-plane MR ($I$$\parallel$$a(b);H$$\parallel$$c$) appears to be small and
smooth through $T_N$, implying that the development of the AF order has
little effect on the in-plane resistivity.
\end{abstract}

}

High-$T_c$ superconductivity (SC) in cuprates occurs as a crossover
phenomenon in the doping range between an antiferromagnetic (AF) insulator
and a Fermi-liquid metal state. While the hole (electron) doping destroys
the long-range AF order in the CuO$_2$ planes, short-range AF correlations
persist well into the superconducting compositions \cite{Fluct}, and thus
it is likely that the interplay of the doped carriers with the AF
correlations underlies the physics of cuprates in a wide range of carrier
concentrations \cite{NAFL,Emery,theor}.

In order to clarify the role of the magnetic interactions in cuprates one
may study the temperature and doping regions which are peculiar for the
spin subsystem. So far, a crossover at a temperature $T^*$ corresponding to
the formation of a pseudogap in the spin and charge excitation spectra has
attracted most attention. An additional decrease in the in-plane
resistivity $\rho_{ab}$ below $T^*$ observed in underdoped
YBa$_2$Cu$_3$O$_{6+x}$ (Y-123) and in YBa$_2$Cu$_4$O$_{8}$ (Y-124) suggests
the possibility that the in-plane transport is determined to a large extent
by the spin scattering \cite{123gap,124gap}. The pseudogap (or spin gap)
was also employed to explain both the activated behavior of the
out-of-plane resistivity $\rho_c(T)$ and the negative out-of-plane
magnetoresistance (MR) in Y-123 and Bi$_2$Sr$_2$CaCu$_2$O$_8$
\cite{ybco1,cMR}.

The vicinity of the N\'{e}el transition is another peculiar region for the
spin subsystem besides the crossover at $T^*$. The dynamic AF correlations
developed in the CuO$_2$ planes evolve into the long-range AF order upon
crossing the N\'{e}el temperature $T_N$ \cite{Fluct}, and one can expect
some singular behavior to show up in the properties governed by the the
magnetic interactions. A magnetotransport study in the vicinity of $T_N$ is
therefore an attractive possibility to clarify the role of spin degrees of
freedom in the peculiar electron transport.

In this Letter, we present a study of the in-plane and out-of-plane MR of
heavily underdoped antiferromagnetic Y-123 crystals, supplemented by
measurements of the Lu-123 crystal used earlier for the study of the phase
diagram \cite{Lav}. We find that the out-of-plane MR undergoes a drastic
change in the vicinity of the N\'{e}el temperature, increasing by about two
orders of magnitude with a transition into the AF state. At the same time,
quite unexpectedly, no feature associated with the AF ordering was observed
in the transverse in-plane MR [$I$$\parallel$$a(b);H$$\parallel$$c$].
Therefore, the development of the AF correlations and the formation of the
long-range N\'{e}el order have a profound influence only on the charge
transport between the CuO$_2$ planes, leaving the in-plane transport
unchanged. Moreover, we find that the longitudinal out-of-plane MR
($H$$\parallel$$c$) is apparently governed by the AF fluctuations even in
the temperature range {\it above} $T_N$, indicating that the spin
fluctuations are playing a major role in the out-of-plane transport
regardless of the presence of the N\'{e}el order.

The high-quality YBa$_2$Cu$_3$O$_{6+x}$ single crystals are grown by the
flux method in Y$_2$O$_3$ crucibles to avoid incorporation of impurities,
and their oxygen content is reduced by subsequent high-temperature
annealing. While the $c$-axis resistivity is easily measured in these
samples owing to the high anisotropy, special care is paid to measure
$\rho_{ab}$ reliably; samples with a length/thickness ratio $\geq$ 100-150
are used and the current contacts are carefully placed to cover the crystal
side surfaces. The MR measurements are performed either by sweeping
temperature (controlled by a Cernox resistance sensor) in constant magnetic
fields up to 16 T, or by sweeping the field at a fixed temperature
stabilized by a capacitance sensor to an accuracy of about 1 mK. The latter
method allows measurements of $\Delta\rho/\rho$ as small as $10^{-5}$ at 10
T.

It was recently reported that heavily underdoped RBa$_2$Cu$_3$O$_{6+x}$
(R=Tm, Lu) exhibits a maximum in $\rho_c(T)$, which originates from a
competition between two distinct mechanisms contributing to the interplane
transport \cite{Lav}. The long-range AF ordering brings about an additional
peculiarity; an abrupt increase in $\rho_c$ occurs upon cooling through
$T_N$, which is followed by a resistivity divergence at lower $T$
\cite{Lav}. We confirm essentially the same behavior in
YBa$_2$Cu$_3$O$_{6+x}$, and in Fig.~1 we present a set of $\rho_c(T)$
curves obtained for the same Y-123 single crystal at slightly different
oxygen contents in the AF region. The rise in $\rho_c$ induced by the AF
transition becomes more and more evident as $T_N$ is lowered; for high
$T_N$, a derivative plot helps to highlight the anomaly and to evaluate
$T_N$ (see inset to Fig.~1). It is worth noting that $T_N$ is extremely
sensitive to the hole doping, as can be seen in Fig.~1, and the width of
the AF transition observed here (10-15 K) is almost the smallest achievable
value, indicating the high quality of our crystals.
\begin{figure}[t]
\leavevmode
\vspace{-14pt}
\leftskip-5pt
\epsfxsize=1.2\columnwidth
\epsffile{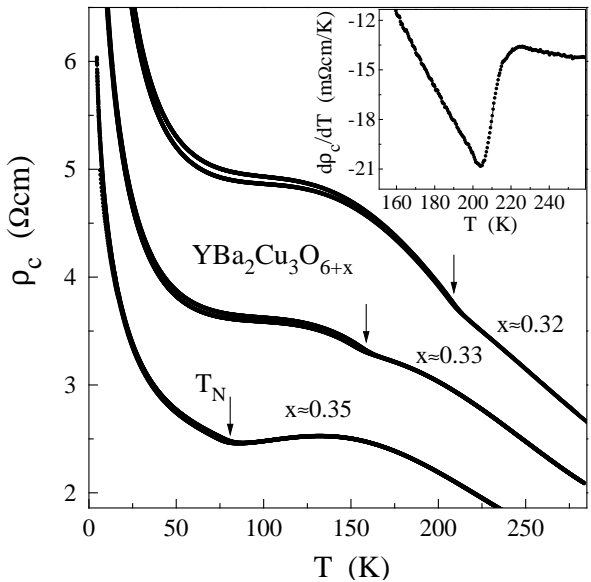}
\vspace{1pt}
\caption{$\rho_{c}(T)$ of an Y-123 single crystal at three different
oxygen contents. Curves for $H$=0 and 16 T [$H$$\parallel$$a(b)$] are
shown. The kink in the resistivity curves marks the AF transition. Inset:
The derivative $d\rho_{c}/dT$ near $T_N$ for $x$$\approx$0.32 ($H$=0).}
\label{fig1}
\end{figure}

Figure 2 demonstrates the unusual behavior of the out-of-plane MR in the
heavily underdoped Y-123, where a step-like increase in
$\Delta\rho_c/\rho_c$ is observed upon cooling through $T_N$. Except for a
small difference in the MR step width, which is obviously related to the
width of the AF transition, this striking feature is very reproducible
within a set of Y-123 crystals and in the Lu-123 crystal. One would expect
that the N\'{e}el transition in Y-123, like other phase transitions
associated with the magnetic subsystem, is considerably affected by the
application of magnetic fields. If a strong magnetic field suppresses AF
order and lowers $T_N$, the out-of-plane MR should then become negative,
because $\rho_c$ is enhanced below $T_N$. What we have found is opposite to
these naive expectations; $T_N$ appears to be quite insensitive to the
magnetic field and the out-of-plane MR, surprisingly, is {\it positive}.
\begin{figure}[t]
\leavevmode
\vspace{-10pt}
\leftskip5pt
\epsfxsize=1.1\columnwidth
\epsffile{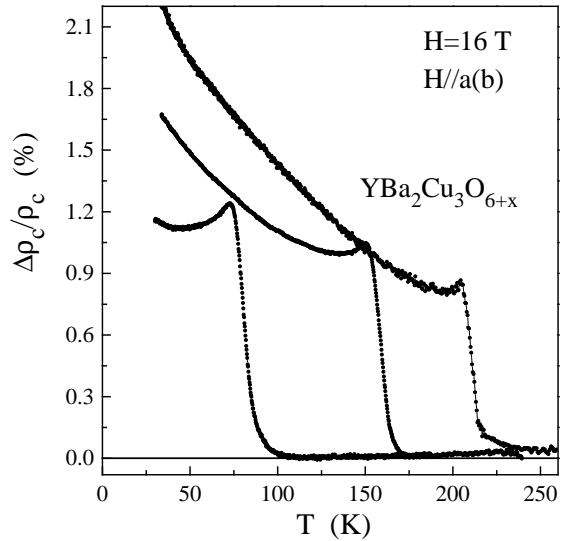}
\vspace{2pt}
\caption{$T$-dependences of the transverse out-of-plane MR
[$H$$\parallel$$a(b)$] at 16 T (same $x$ values as in Fig.~1).}
\label{fig2}
\end{figure}

Upon tilting the magnetic field (inset to Fig.~3), we observe a
considerable anisotropy in the out-of-plane MR, $\Delta\rho_c/\rho_c$,
which becomes largest for the transverse geometry
$H$$\perp$$I$$\parallel$$c$. Besides the difference in the magnitude, the
$T$ dependence of the out-of-plane MR is remarkably different between the
$H$$\parallel$$ab$ and the $H$$\parallel$$c$ geometries; while
$\Delta\rho_c/\rho_c$ for $H$$\parallel$$ab$ keeps growing below $T_N$
(Fig.~2) after it shows a small peak, $\Delta\rho_c/\rho_c$ for
$H$$\parallel$$c$ gradually diminishes below $T_N$ after it shows a
pronounced peak (Fig.~3). Therefore, the magnetoresistance becomes more
anisotropic as the temperature is lowered below $T_N$.

\begin{figure}[t]
\leavevmode
\vspace{-4pt}
\epsfxsize=1.18\columnwidth
\epsffile{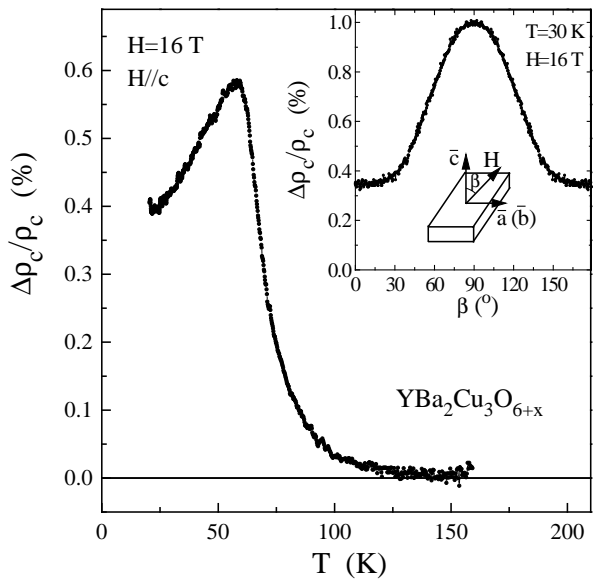}
\vspace{2pt}
\caption{$T$-dependences of the longitudinal out-of-plane
MR ($H$$\parallel$$c$) at 16 T. Inset: Angular dependence of the
out-of-plane MR measured at 30 K for a Y-123 crystal with $T_N$=108 K
(measurements performed within the 100$^{\circ}$ range and extended for
other angles).}
\label{fig3}
\end{figure}

To obtain an idea about the mechanisms which couple mobile carriers with
the AF order, it is helpful to look also at the in-plane transport. For the
oxygen contents under study the in-plane resistivity demonstrates a
crossover behavior, passing through a minimum at $T=50-60$ K [Fig.~4(a)].
To our surprise, the in-plane MR $\Delta\rho_{ab}/\rho_{ab}$, as well as
$\rho_{ab}$ itself, is always smooth in the vicinity of $T_N$ and we do not
find any anomaly which can be associated with the N\'{e}el transition. For
the transverse in-plane MR [$H$$\parallel$$c$; $I$$\parallel$$a(b)$], a
small magnitude of the MR and a possible admixture of the asymmetric Hall
component ($\propto H$) in raw data required more precise measurements with
field sweeping to be performed; the resulting field dependences of
$\rho_{ab}$ are presented in Fig.~4(b). The $T$ dependence of
$\Delta\rho_{ab}/\rho_{ab}$ is shown in Fig.~4(c), where the in-plane MR
reveals no correlation with the AF transition even at this sensitivity
level ($\sim$10$^{-5}$). Instead it is rather small and remains almost
constant down to the temperature region where $\rho_{ab}$ acquires
localizing behavior [Fig.~4(a)] and $\Delta\rho_{ab}/\rho_{ab}$ changes its
sign. The N\'{e}el transition and the corresponding changes in the
spin-excitation spectrum have therefore no effect on the charge transport
within the CuO$_2$ planes.
\begin{figure}[t]
\leavevmode
\vspace{-38pt}
\leftskip-12pt
\epsfxsize=1.12\columnwidth
\epsffile{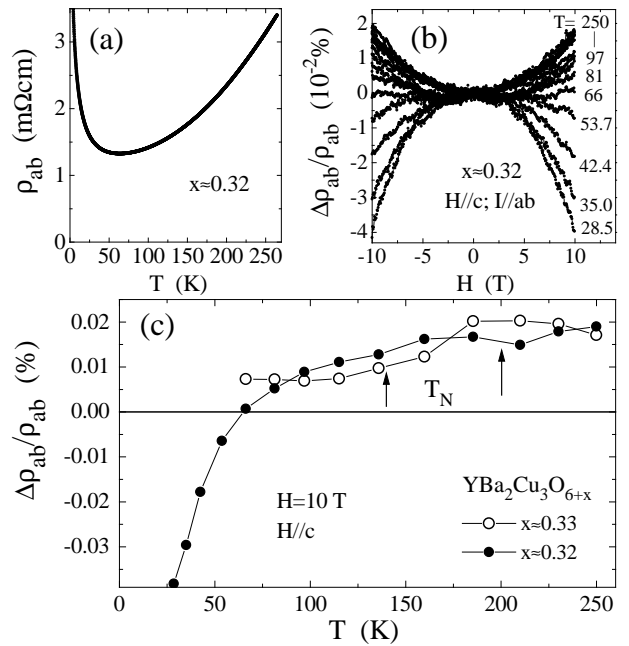}
\vspace{-82pt}
\caption{(a) $\rho_{ab}(T)$ of an Y-123 single crystal. (b) $H$-dependences
of the transverse in-plane MR. (c) $T$-dependences of the MR at 10 T for
two $x$ values.}
\label{fig4}
\end{figure}

An intriguing issue is whether the out-of-plane transport is sensitive
exclusively to the long-range order arising below $T_N$. If the short-range
AF correlations above $T_N$ also contribute to $\rho_c$ and its MR, we may
expect that the AF fluctuations play an essential role not only in the AF
compositions but also in the SC compositions. Apparently, the more precise
field-sweeping technique should be employed to investigate the behavior
above $T_N$, where the MR becomes very small. Besides, below $T_N$ such
measurements allow us to single out the main $\gamma H^2$ term from
possible additional contributions to the MR \cite{comment}.
\begin{figure}[ht]
\vspace{-5pt}
\leftskip-9pt
\epsfxsize=1.10\columnwidth
\epsffile{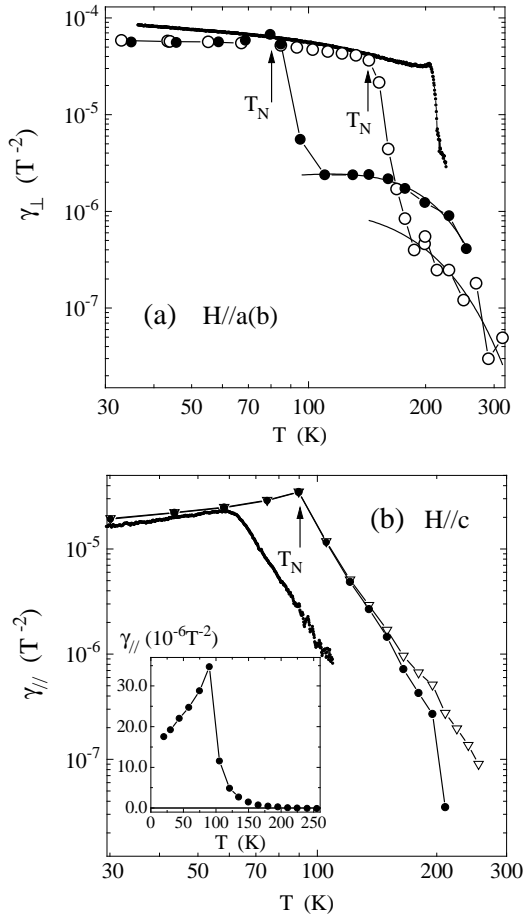}
\caption{(a) $T$-dependences of the $\gamma_{\perp}H^2$ term in the
transverse out-of-plane MR for Y-123 (open circles) and Lu-123 (solid
circles).  The arrows indicate $T_N$. (b) $T$-dependence of the
$\gamma_{\parallel}H^2$ term ($H$$\parallel$$c$$\parallel$$I$) for Lu-123
(solid circles). The open triangles illustrate the data for Lu-123 after
subtracting the weak negative background. Inset: The Lu-123 data on a
linear scale. In both (a) and (b), the MR data for Y-123 for sweeping of
the temperature are shown (dots) for comparison.}
\label{fig5}
\end{figure}
The $T$ dependences of $\gamma_{\perp} H^2$ and $\gamma_{\parallel} H^2$
components of $\Delta\rho_c/\rho_c$ (for $H$$\perp$$c$ and
$H$$\parallel$$c$, respectively) presented in Fig.~5 depict the qualitative
difference in the MR behavior for the two directions of the magnetic field.
For $H$$\parallel$$ab$ [Fig.~5(a)], the out-of-plane MR changes at $T_N$ in
a step-like manner by up to two orders of magnitude, where the width of the
step is the same as the width of the AF transition itself. The step
separates regions below and above $T_N$ with relatively weak dependence of
the MR on temperature. This behavior implies that the sensitivity to the
magnetic field appears abruptly with the onset of the long-range AF order.
On the other hand, for $H$$\parallel$$c$ [Fig.~5(b)] we observe a MR peak
at $T_N$, which is accompanied by a tail spreading to far above $T_N$. The
MR as a function of temperature has no discontinuity at $T_N$ and one can
infer from Fig.~5(b) that $\Delta\rho_c/\rho_c$ grows as $T^{-k}$ with
lowering $T$ until the N\'{e}el transition interrupts this tendency.
However, the right-hand side of the MR peak for $H$$\parallel$$c$ (the
$T^{-k}$ behavior) apparently shifts with $T_N$ when the $x$ is changed,
which indicates its relation to the AF ordering. Therefore, we can conclude
that a mechanism associated with the AF fluctuations dominates the
out-of-plane MR in a wide temperature range above $T_N$ as well.  This
observation clearly demonstrates that the short-range AF correlations play
an essential role in the interplane transport. At high $T$, the
longitudinal MR turns out to be weakly [$<$(1.5-2.5)$\times$10$^{-5}$ at 10
T] negative [Fig.~5(b)], which is reminiscent of the large negative MR
observed in moderately underdoped Y-123 \cite{cMR}. We note that this weak
negative background has a negligible effect on the MR behavior in the
temperature range up to $2T_N$ [Fig.~5(b) shows how its subtraction
modifies the data] and is not important for the present discussion.

The contrasting behavior of the in-plane and out-of-plane MR indicates that
changes which occur in the spin subsystem at $T_N$ are influential only on
the electron transport between the CuO$_2$ planes and apparently not on the
in-plane one. It is known that the heavily underdoped Y-123 above $T_N$
possesses well-developed dynamic AF correlations in the CuO$_2$ planes
\cite{Fluct} and the N\'{e}el temperature actually corresponds to the
establishment of AF order along the $c$-axis. The symmetry change
accompanying the long-range order and a change in the spin dynamics are the
only two mechanisms for the N\'{e}el transition to influence the electron
transport. In spite of the sharp increase in $\rho_c$ upon cooling through
$T_N$, opening of a gap in the quasiparticle energy spectrum due to the
magnetic superstructure is unlikely, since one can hardly imagine a gap
formation to have no impact on the in-plane transport. On the other hand,
it is possible that the freezing of the spin degrees of freedom below $T_N$
causes an increase in $\rho_c$, if the spin fluctuations assist the
electron hopping between the CuO$_2$ planes. Since an increase in $\rho_c$
also takes place when the magnetic field is applied, one can infer that the
field suppression of the spin fluctuations is likely to be the main source
of the positive out-of-plane MR in our heavily underdoped Y-123. Also, the
dramatic changes in the out-of-plane transport associated with the
evolution of the magnetic state might suggest that it is the spin subsystem
that is responsible for the charge confinement within the CuO$_2$ planes.

Now let us discuss the actual mechanism which gives rise to the peculiar
transport properties observed here. A possible picture to account for the
observed features is the segregation of the doped holes into ``stripes''
which separate AF domains \cite{Emery}. In this picture, the confinement of
the charges into the CuO$_2$ planes is substituted in a sense by the
confinement to the quasi-1D stripes. Since the formation of the stripes
themselves is governed by the magnetic interactions, it is not surprising
that the spin degrees of freedom are playing a dominant role in the hole
confinement and hence in the out-of-plane charge transport. Also, we can
expect the in-plane MR to be very weak in this picture; the orbital MR term
is irrelevant for the carriers moving along quasi-1D stripes, since the
magnetic field cannot bend their trajectories. The spin-charge separation
(naturally expected for 1D stripes \cite{Emery}) and the spin gap formed in
both the stripes and their AF environment \cite{Emery} imply that the
spin-dependent scattering for the charge transport along stripes should not
be large.

In summary, the out-of-plane transport in heavily underdoped
YBa$_2$Cu$_3$O$_{6+x}$ is found to show anomalous magnetoresistance
associated both with the N\'{e}el ordering and with the AF correlations
above $T_N$. The MR behavior gives evidence that the spin fluctuations play
an essential role in the interplane transport and hence suggest that the
charge confinement within the CuO$_2$ planes is also fundamentally related
to the spin degrees of freedom.

We are grateful to L. P. Kozeeva for providing the Lu-123 crystals. A.N.L.
acknowledges the support from JISTEC through an STA fellowship.


\begin{thebibliography}{99}
\bibitem{comment} The field dependence of $\rho_c$ at $T<~T_N$ is not
always trivial; for samples with $T_N$ $\leq$ 50-60 K,
$\Delta\rho_c/\rho_c$ is $\propto H^2$ up to 16 T, but an additional
negative low-field MR term appears to be superimposed on the $\gamma H^2$
behavior as the doping level of the sample is further reduced. This
additional low-field MR feature can be explained by the rearrangement of
the AF domain boundaries (stripes) \cite{ourstr}.
\end{thebibliography}
\end{document}